\documentclass[prb,twocolumn,showpacs,preprintnumbers,amsmath,amssymb]{revtex4}

\usepackage{hyperref}
\usepackage{graphicx}

\newcommand {\be}{\begin{eqnarray}}
\newcommand {\ba}{\begin{align}}
\newcommand {\ee}{\end{eqnarray}}
\DeclareMathAlphabet{\ib}{OML}{cmm}{b}{it}

\begin{document}


\title{Coexistence of phase transitions and hysteresis near BEC} 


\author{M.~M{\"a}nnel}
\email[]{maennel@fh-muenster.de}
\affiliation{M{\"u}nster University of Applied Sciences, Stegerwaldstrasse 39, 48565 Steinfurt, Germany}
\author{K.~Morawetz}
\affiliation{M{\"u}nster University of Applied Sciences, Stegerwaldstrasse 39, 48565 Steinfurt, Germany}
\affiliation{International Institute of Physics (IIP), 
Av.~Odilon Gomes de Lima 1722, 59078-400 Natal, Brazil}
\affiliation{Max Planck Institute for the Physics of Complex
Systems, Noethnitzer Str. 38, 01187 Dresden, Germany}
\author{P.~Lipavsk\'y}
\affiliation{Faculty of Mathematics and Physics, Charles University, Ke Karlovu 3, 12116 Prague 2, Czech Republic}


\date{\today}

\begin{abstract}
Multiple phases occurring in a Bose gas with finite-range interaction are investigated. In the vicinity of the onset of Bose-Einstein condensation (BEC) the chemical potential and the pressure show a van-der-Waals like behavior indicating a first-order phase transition although there is no long-range attraction. 
Furthermore the equation of state becomes multivalued near the BEC transition. For a Hartree-Fock or Popov (Hartree-Fock-Bogoliubov) approximation such a multivalued region can be
avoided by the Maxwell construction. For sufficiently weak interaction the multivalued region can also be removed using a many-body \mbox{T-matrix} approximation. However, for strong interactions there remains a multivalued region even for the \mbox{T-matrix} approximation and after the Maxwell construction, what is interpreted as a density hysteresis. This unified treatment of normal and condensed phases becomes possible due to the recently found scheme to eliminate self-interaction in the \mbox{T-matrix} approximation, which allows to calculate properties below and above the critical temperature.
\end{abstract}

\maketitle


\section{Introduction}
 When Einstein predicted BEC \cite{E25} for an ideal gas of bosons extending a paper by Bose, it was not foreseeable that it would need 70 years before experimental verification, which was performed in $^{87}$Rb by \cite{AEMWC95}, in $^7$Li by \cite{BSTH95}, and in $^{23}$Na by \cite{DMADDKK95} at temperatures between $0.1$ and \mbox{$2$ $\mu$K}. These measurements have encouraged an enormous theoretical activity among which the problem to account adequately for correlations is still unsettled. Specific interesting consequences of correlations are the change of condensation temperature  \cite{A04,BBHLV01,HGL99,BR02,KNP04,HLK04,MMS07}, the occurrence of further phase transitions and even the change of the nature of the BEC transition itself. Since this is not the place to give credit to all these important activities, we want to focus on the single questions of possible phase transitions due to correlations. Even the BEC is sometimes viewed as a first-order phase transition \cite{H85} which seems to be doubtful when attributing a phase transition to interactions and the BEC appearing already in ideal gases.

Multiple phase transitions have been reported e.g. in \cite{SPR98} where the influence of BEC to the liquid-gas phase transition has been calculated. We will follow this path and explore the coexistence and mutual influence of a phase transition and the BEC. Since now there is a consistent scheme available which allows to describe the situation in and out of the BEC by a common theoretical object, the multiple-scattering-corrected T-matrix, we are in the position to investigate the mutual influence of phase transitions and the BEC, due to interactions. This leads to the expectation that interactions and correlations are a proper tool to tune the BEC parameter since they can be controlled fairly well e.g. by Feshbach resonances \cite{IASMSK98,CFHAV98,RCBGCW98}. 

Strongly correlated systems are 
connected with a highly nonlinear density dependence of the thermodynamic
quantities. Interestingly such nonlinearities can lead to
hysteresis behavior. Besides the known magnetic-field hysteresis there are a
number of examples observed in other fields. Optical bistable systems  have been
reported to show a time-hysteresis in the response due to a nonlinear density
dependence \cite{LFCGY94}. A pressure induced thermal hysteresis in Kondo
lattice systems has been found \cite{PSLFT05} and even in plasma discharge
systems a density-driven hysteresis is reported \cite{JWZW09}. A density
hysteresis driven by pressure can be found in spin-crossover compounds 
due to elastic stresses \cite{SCHSE11}. Near the BEC of a quantum
spin system a peak in sound attenuation was observed \cite{SLBOT03} 
and attributed to the
hysteresis in magnetic field, which indicates a first-order phase-transition.
Here in this paper we show that in strongly correlated Bose systems near BEC 
a density hysteresis appears.

The outline of the paper is as follows. In the next section~\ref{sectn2} we explain the main set of equations of the \mbox{T-matrix} approach with multiple scattering corrections and how known approximations appear. 
The condensed and non-condensed phase can be described in this way on the same theoretical footing. In section~\ref{sectn3} we discuss the solution in terms of the equation of state. We show that the appearing artifact of a multivalued region is reduced with increasing level of approximation \cite{HB03,PRS04} and can be avoided by the Maxwell construction. Furthermore we investigate how the Bose condensate behaves during a first-order phase-transition. For the strong interacting limit we report a phase of back bending of density with increasing chemical potential which indicates an anomalous rearrangement and which we interpret as density hysteresis. The comparison with other approaches and with experiments is discussed in section~\ref{sectn4}. Section~\ref{sectn5} finally contains the summary and conclusions.   

\section{The T-matrix approximation}
\label{sectn2}
We will present a consistent treatment of interactions and condensation in a unified manner with the help of the corrected multiple-scattering T-matrix which yields a non-perturbative description of strong correlations beyond the mean field.  Our starting point is a homogeneous gas of interacting Bosons with mass $m$, temperature $T$ and particle density $n$. The temperature $T$ scales in energy units so that Boltzmann's constant $k_{\rm B}$ can be omitted. The Hamiltonian has the structure
\be
\hat H\!=\!\sum_{\ib k}\!\frac{\hbar^2k^2}{2m}\hat a_{\ib k}^\dagger\hat a_{\ib k} +\frac{\lambda}{2\Omega}\!\sum_{\ib q,\ib p,\ib k}\! \hat a_{\ib p}^\dagger\hat a_{\ib q-\ib p}^\dagger \hspace*{.05cm}g_{\ib p-\frac{\ib q}{2}}\hspace*{.05cm}g_{\ib k-\frac{\ib q}{2}}\hspace*{.05cm}\hat a_{\ib k}\hat a_{\ib q-\ib k},\nonumber\\
\ee
where $\hat a_{\ib k}^\dagger$ $\left(\hat a_{\ib k}\right)$ creates (annihilates) a particle with momentum $\ib k$. The volume of the system $\Omega$ is considered in the thermodynamic limit $\Omega\to\infty$. The interaction is characterized by the strength $\lambda$ and the Yamaguchi form factors $g_{\ib p}=\left(1+p^2/\gamma^2\right)^{-1}$ \cite{Y54}. The latter yields a soft momentum cut-off to avoid an ultraviolet divergence. The parameter $\gamma$ is related to the range of the interaction. 
$\lambda$ is positive for repulsive interaction. 

\subsection{Condensed phase}
In the BEC phase a fraction of particles is condensed, with a condensate density $n_{\bf 0}$. 

We use a scheme to eliminate self-interaction in the \mbox{T-matrix} approximation, to calculate properties below the critical temperature \cite{
L08,M10,MML10,SLMMM11}. The Green function \cite{SG98,M11} for particles with momentum $\ib q$ 
and Matsubara frequency $iz_\nu=2\pi\nu T$, $\nu\in\mathbb Z$, is 
\be
G(\ib q,iz_\nu)=\frac{iz_\nu+\epsilon_{\ib q}}{iz_\nu^2-\epsilon_{\ib q}^2+n_{\bf 0}^2{\cal T}^2(\ib q)}=\frac{iz_\nu+\epsilon_{\ib q}}{iz_\nu^2-E_{\ib q}^2}
\label{2}
\ee
where the interactions between the particles are considered in a ladder-summation of diagrams resulting into the many-body T-matrix
\be
{\cal T}(\ib q)
=\lambda g_{\ib q}\left(1+\lambda\int\frac{d^3k}{(2\pi)^3}\frac{g^2_{\ib k}}{2E_{\ib k}}(1+2f_{\rm B}(E_{\ib k}))\right)^{-1}\nonumber\\
\label{tm}
\ee
and $f_{\rm B}(\epsilon)=1/\left(e^{\epsilon/T}-1\right)$ is the Bose distribution function.  
The quasi-particle dispersion is given by the poles of the Green function. In the normal phase, for $n_{\bf 0}=0$, the dispersion would be
\be
\epsilon_{\ib q}=\frac{\hbar^2q^2}{2m}-\mu+2n{\cal T}({\bf 0}).
\label{disp0}
\ee
In the condensed phase the chemical potential $\mu$ satisfies the Hugenholtz-Pines \cite{HP59} relation 
\mbox{$
\mu=2n{\cal T}({\bf 0})-n_{\bf 0}{\cal T}({\bf 0})
$} and the Green function yields the generalized Bogoliubov dispersion 
\be
E_{\ib q}=\sqrt{\left(\frac{\hbar^2q^2}{2m}+n_{\bf 0}{\cal T}({\bf 0})\right)^2-n_{\bf 0}^2{\cal T}^2(\ib q)}.
\label{disp}
\ee
With the particle density
\be
n
&=&-\frac{T}{\Omega}\sum\limits_{\ib k,\nu} G({\ib k},iz_\nu)=n_{\bf 0}+\!\int\!\frac{d^3k}{(2\pi)^3}\left(1+2v_{\ib k}^2\right)f_{\rm B}(E_{\ib k})
\nonumber\\
&&+\int\!\frac{d^3k}{(2\pi)^3}v_{\ib k}^2,
\label{n}
\ee
the set of \mbox{equations~(\ref{tm}-\ref{n})} is closed. The depletion of the condensate at $T=0$ is described by \mbox{$v_{\ib k}^2=\left(\epsilon_{\ib k}-E_{\ib k}\right)/2E_{\ib k}$}. Also the expectation value of the total energy density can be calculated from the Green function \cite{KB88}
\be
u&=&\frac{\langle\hat H\rangle}{\Omega}=-\frac{T}{\Omega}\sum\limits_{\ib k,\nu}\frac{1}{2}\left(iz_\nu+\mu+\frac{\hbar^2k^2}{2m}\right) G({\ib k},iz_\nu)\nonumber\\
&=&\underbrace{\int\!\frac{d^3k}{(2\pi)^3}E_{\ib k}f_{\rm B}(E_{\ib k})}_{u_{\rm qp}}\!+\! \underbrace{{\cal T}({\bf 0})\left(n^2\!-\!nn_{\bf 0}\!+\!\frac{1}{2}n_{\bf 0}^2\right)}_{u_{\rm mf}}\nonumber\\
&&\underbrace{-\int\!\frac{d^3k}{(2\pi)^3}E_{\ib k}v^2_{\ib k}}_{u_{\rm cor}}
\!+\!\underbrace{\int\!\frac{d^3k}{(2\pi)^3}\frac{n_{\bf 0}^2{\cal T}^2(\ib k)}{4E_{\ib k}}(1\!+\!2f_{\rm B}(E_{\ib k}))}_{u_{\rm 2p}}.\nonumber\\
\label{7}
\ee

Lets inspect different levels of approximation and the corresponding contributions to this energy density. The mean-field-like approximation ${\cal T}(\ib q)\approx\lambda$ together with\\ \mbox{$
E_{\ib q}=\epsilon_{\ib q}\approx\hbar^2q^2/2m+n_{\bf 0}\lambda
$} establishes the 
Hartree-Fock approximation as proposed by Huang et al. \cite{HY57,HYL57} and in (\ref{7}) only the contribution of quasi particles $u_{\rm qp}$ and the mean field term $u_{\rm mf}$ survive leading to
\be
u=\int\frac{d^3k}{(2\pi)^3}\frac{\hbar^2k^2}{2m}f_{\rm B}(\epsilon_{\ib k})+\lambda\left(n^2- \frac{1}{2}n_{\bf 0}^2\right).
\ee
This energy density shows, that, in addition to statistics, BEC is also energetically favored, since a finite condensate density $n_{\bf 0}$ lowers the interaction energy. This phenomenon is called ''attraction in momentum space'' \cite{H642,L01}.

Approximating only ${\cal T}(\ib q)\approx\lambda$ provides the Hartree-Fock-Bogoliubov or Popov approximation, with the typical Bogoliubov dispersion 
\be
E_{\ib q}=\sqrt{\left(\frac{\hbar^2q^2}{2m}+n_{\bf 0}\lambda\right)^2-n_{\bf 0}^2\lambda^2}.
\ee 
Within this approximation a further contribution of the energy density (\ref{7}) remains besides the quasi particle and the mean field term, a correlation term $u_{\rm cor}$, which favors a finite depletion \cite{B47,PS04}. It has to be noted that the original Bogoliubov approximation corresponds to an additional approximation of the chemical potential $\mu\approx n_{\bf 0}\lambda$.

For the T-matrix approximation there appears a fourth contribution to the energy density (\ref{7}),
\be
u_{\rm 2p}\!=\!- \frac{1}{2}\!\int\!\frac{d^3k}{(2\pi)^3}\frac{d^3q}{(2\pi)^3}\lambda g_{\ib q}g_{\ib k}C_{\ib q}C_{\ib k}-\frac{n_{\bf 0}}{2}\!\int\!\frac{d^3k}{(2\pi)^3}\lambda g_{\ib k}C_{\ib k}
\nonumber\\
\ee
which is a two-particle term, that can be expressed by the anomalous expectation value of a pair of particles
\be
C_{\ib k}=\langle\hat a_{\ib k}\hat a_{-\ib k}\rangle=-\frac{n_{\bf 0}{\cal T}(\ib k)}{2E_{\ib k}}(1+2f_{\rm B}(E_{\ib k})).
\ee
Please note that this two-particle term appears as a consequence of the theory \cite{M10} here and has not been assumed ad-hoc as done in most approaches postulating anomalous functions. 
A very similar term can be found in the Bardeen-Cooper-Schrieffer (BCS) approximation \cite{PS04} where it describes the contribution of Cooper pairs.

%
%
%
%

\subsection{Normal phase}
Due to the use of the corrected multiple-scattering \mbox{T-matrix} approximation, i.e., due to the elimination of self-interaction, equations (\ref{2}), (\ref{tm}) and (\ref{n}) are valid in the normal phase as well and yield the same level of approximation as in the BEC phase \cite{L08,SLMMM11,M11}. For $n_{\bf 0}=0$, the Green function simplifies to
\be
G(\ib q,iz_\nu)=\frac{1}{iz_\nu-\epsilon_{\ib q}},
\ee
with the dispersion $E_\ib q=\epsilon_{\ib q}$ according to (\ref{disp0}). The corresponding particle density is
\be
n=\int\frac{d^3k}{(2\pi)^3}f_{\rm B}(\epsilon_{\ib k}),
\ee
and the energy density is
\be
u=\int\frac{d^3k}{(2\pi)^3}\frac{\hbar^2k^2}{2m}f_{\rm B}(\epsilon_{\ib k})+{\cal T}({\bf 0})n^2.
\ee
In this approximation each quasi particle simply feels a mean field $2n{\cal T}({\bf 0})$.

In the normal phase the Popov approximation is identical to the Hartree-Fock approximation, with a mean field $2n\lambda$. While the Bogoliubov approximation yields an ideal Bose gas.

\section{Equation of state}
\label{sectn3}
\subsection{Chemical potential}

\begin{figure}[t]
\includegraphics[width=8cm]{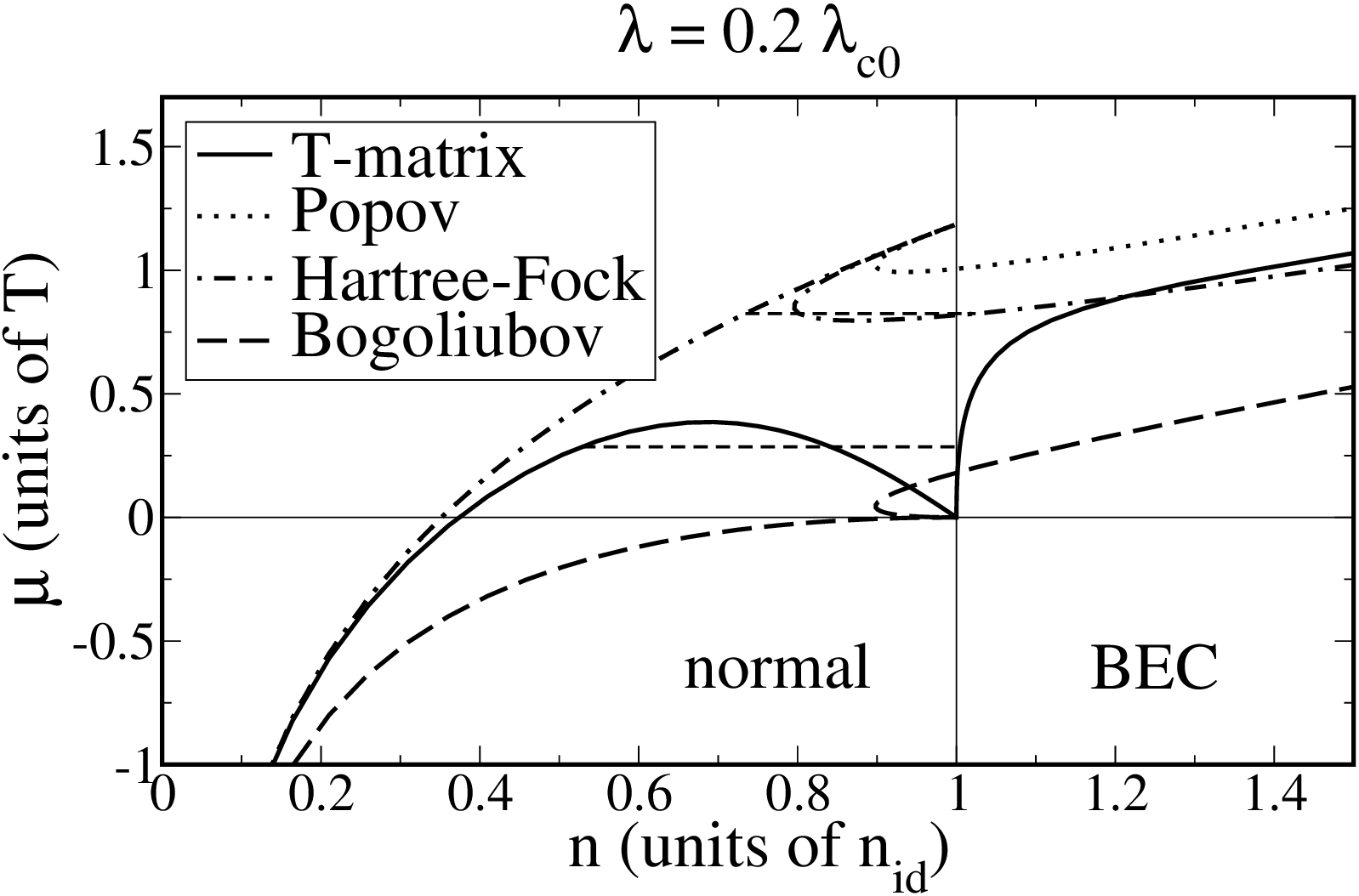}
\caption{Chemical potential in Bogoliubov, Hartree-Fock, Popov and T-matrix approximation for weak repulsive interaction, the horizontal broken lines correspond to the Maxwell construction, ${n_{\rm id}\approx
2.61/\Lambda_{\rm dB}^{3}}$ is the ideal critical density for Bose condensation, the constant quantity is given above the diagram, 
${\Lambda_{\rm dB}=
\hbar\sqrt{2\pi/mT}}$ is the thermal de Broglie wavelength, ${\lambda_{\rm c0}=
4\pi\hbar^2\Lambda_{\rm dB}/m\sqrt{\pi}}$ and $\gamma=2\sqrt{\pi}/\Lambda_{\rm dB}$.
\label{cnth}}
\end{figure}

\begin{figure}[t]
\includegraphics[width=8cm]{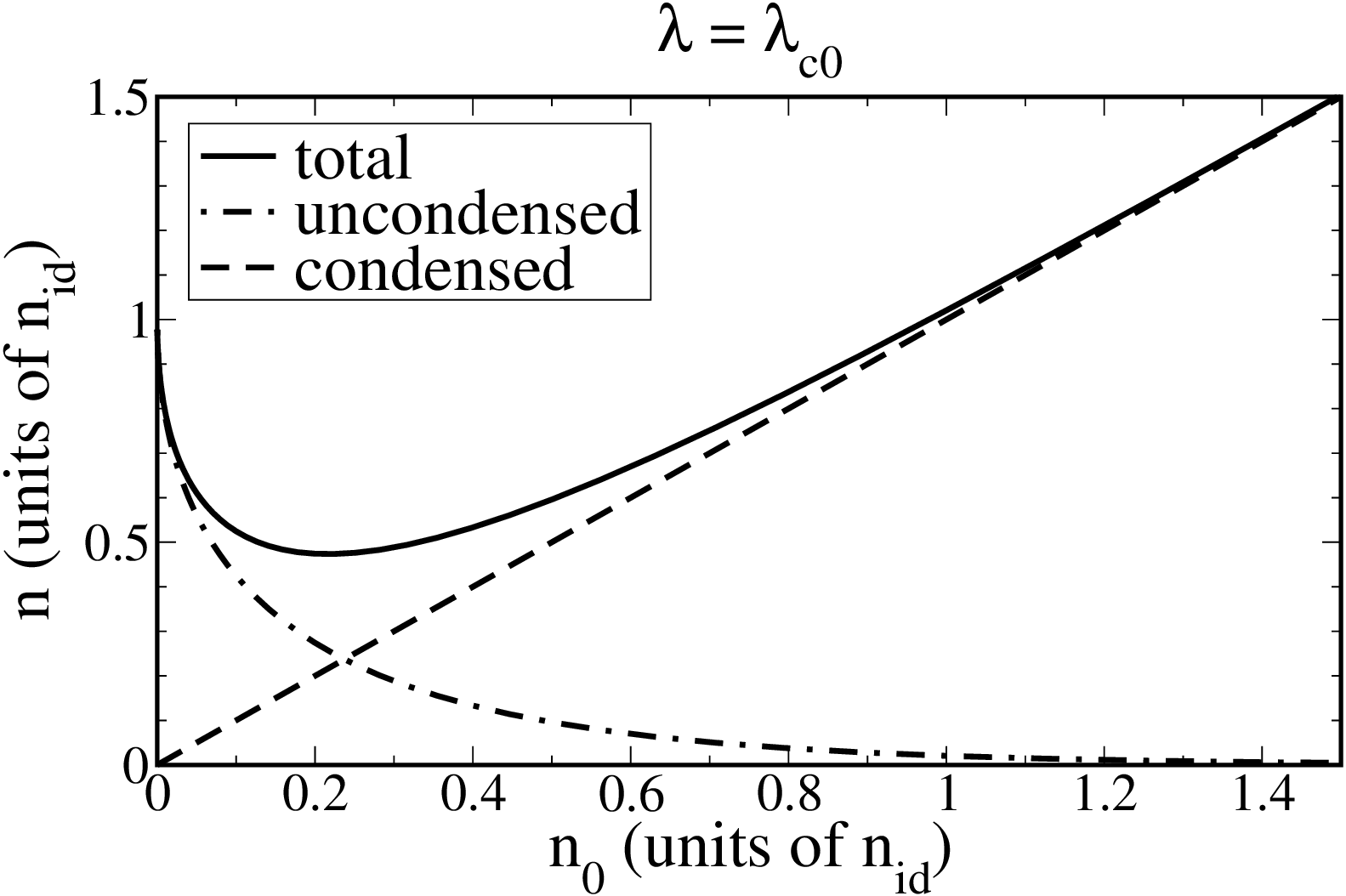}
\caption{Contributions of condensed and non-condensed particles to the total particle density $n$ in 
Hartree-Fock approximation.
\label{nn0huang3}}
\end{figure}

\begin{figure}[t]
\includegraphics[width=8cm]{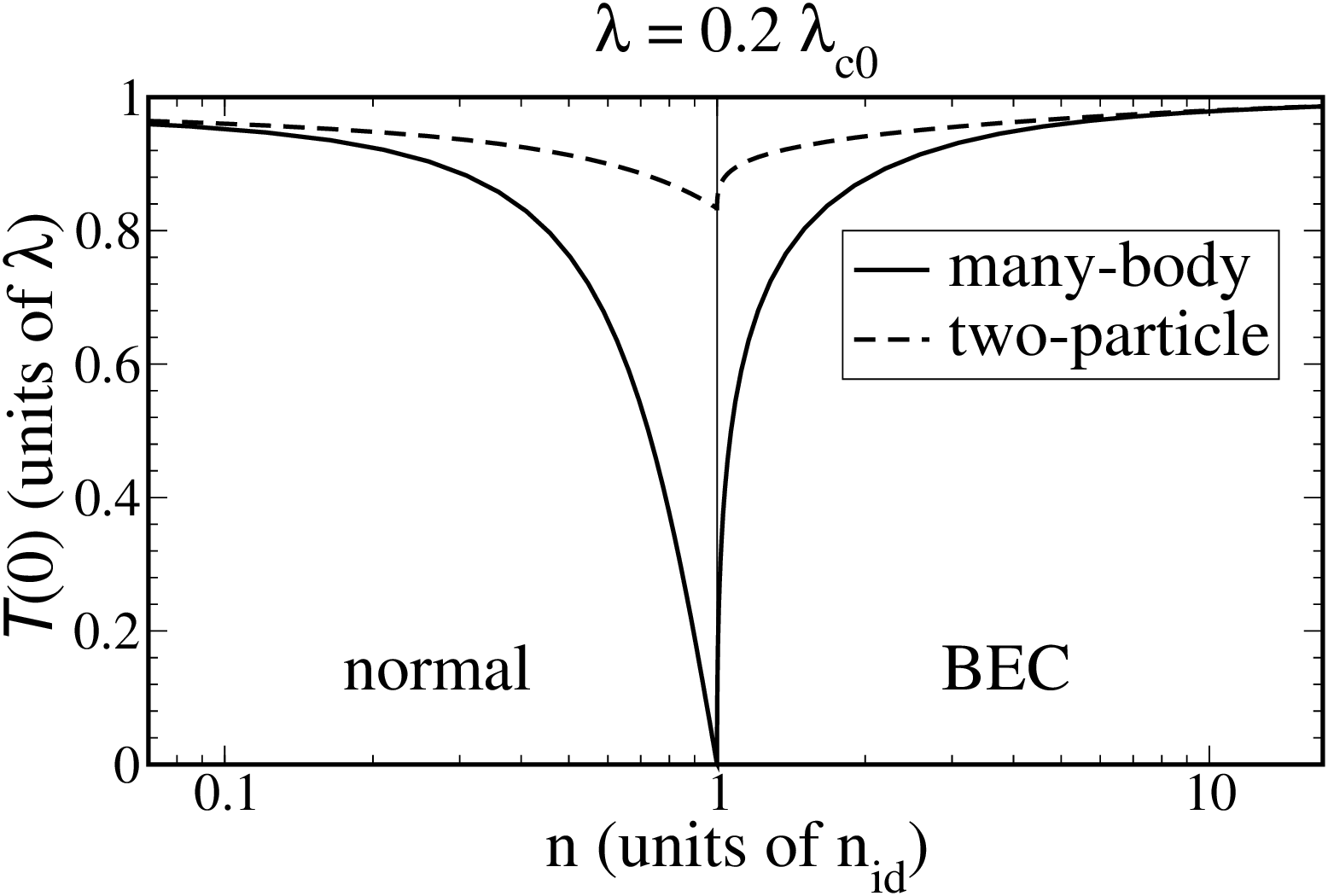}
\caption{Many-body and two-particle T-matrix for zero momentum and energy in T-matrix approximation.
\label{tvsn2}}
\end{figure}

From the set of equations we calculate the chemical potential $\mu$ for different particle densities $n$, as shown in Fig.~\ref{cnth}.
The 
Hartree-Fock approximation, i.e. the dashed-dotted line, shows a multivalued region near the onset of BEC, where several solutions of the equation of state coexist. The origin of this unphysical behavior seems to be an overestimation of the attraction in momentum space,  
which favors a high condensate fraction. Fig.~\ref{nn0huang3} illustrates a rapid drop of the density of non-condensed particles after the onset of BEC. That is due to the attraction in momentum space and leads to a temporary drop of the total density, which is responsible for the back-and-forth behavior of the chemical potential in Fig.~\ref{cnth}. Furthermore Fig.~\ref{cnth} shows that there is also a temporary drop of the chemical potential in  
Hartree-Fock approximation after BEC has set in, which indicates an instability of the gas and a first-order phase-transition. It has to be emphasized however, that this instability has its origin not in the attractive part of the interaction potential but in the BEC and the attraction in momentum space. During the first-order phase-transition there is a coexistence of a high- and a low-density phase and according to Gibb's phase rule there is only one free parameter which has to be constant in order to keep the temperature fixed. Therefore, all intensive parameters of the two phases are constant during the phase transition, especially pressure and chemical potential. In equilibrium, the pressure and the chemical potential have to be equal for both phases and can be obtained via the Maxwell construction, illustrated in Fig.~\ref{cnth} by the horizontal broken line. As the system follows the curve of constant pressure and chemical potential the unphysical multivalued region is avoided.

As illustrated by the broken curve in Fig.~\ref{cnth}, the chemical potential in the Bogoliubov approximation shows an unphysical region as well. However, in this approximation the Maxwell construction is not possible. Since the Bogoliubov approximation was developed to describe the system near $T=0$, it fails near the BEC transition. The approximation can be improved by including the Hartree-Fock mean field, leading to the Popov approximation. Although the unphysical region remains, the Maxwell construction becomes possible. Compared to the 
Hartree-Fock approximation the width of the unphysical region is reduced with no qualitative change. Therefore in the following it is sufficient to compare only the 
Hartree-Fock approximation with our T-matrix approximation.

If the repulsive interaction is weak, i.e., \mbox{$\lambda<0.23$~$\lambda_{\rm c0}$}, the chemical potential in T-matrix approximation shows no unphysical region. Nevertheless there is still an instability of the gas, i.e., the chemical potential drops down to zero at the onset of BEC, shown by the full line in Fig.~\ref{cnth}. The reason for the vanishing of the chemical potential is the phenomenon that the many-body T-matrix for zero momentum and energy ${\cal T}({\bf 0})$ vanishes at the critical point \cite{BS97,SG98}, as illustrated by the full line in Fig.~\ref{tvsn2}. 

As shown in Fig.~\ref{tvsn2} the vanishing of the many-body \mbox{T-matrix} is clearly a medium effect since the two-particle \mbox{T-matrix} which does not include medium contributions stays finite at the critical density for BEC. The two-particle \mbox{T-matrix}, i.e., the broken line, can be obtained from (\ref{tm}) by omitting the Bose function. It seems that in the vicinity of the onset of BEC the repulsive interaction is compensated by the Bose enhancement, leading to the drop of the chemical potential. Again there is a first-order phase-transition due to this instability and the Maxwell construction yields the critical chemical potential and pressure. However, the drop of the chemical potential and the corresponding first-order phase-transition might as well be an artifact, of omitting the momentum dependence of the \mbox{T-matrix} in the self energy, leading to the dispersion (\ref{disp0}) \cite{BS97,HB03,PRS04}.

\begin{figure}[t]
\includegraphics[width=8cm]{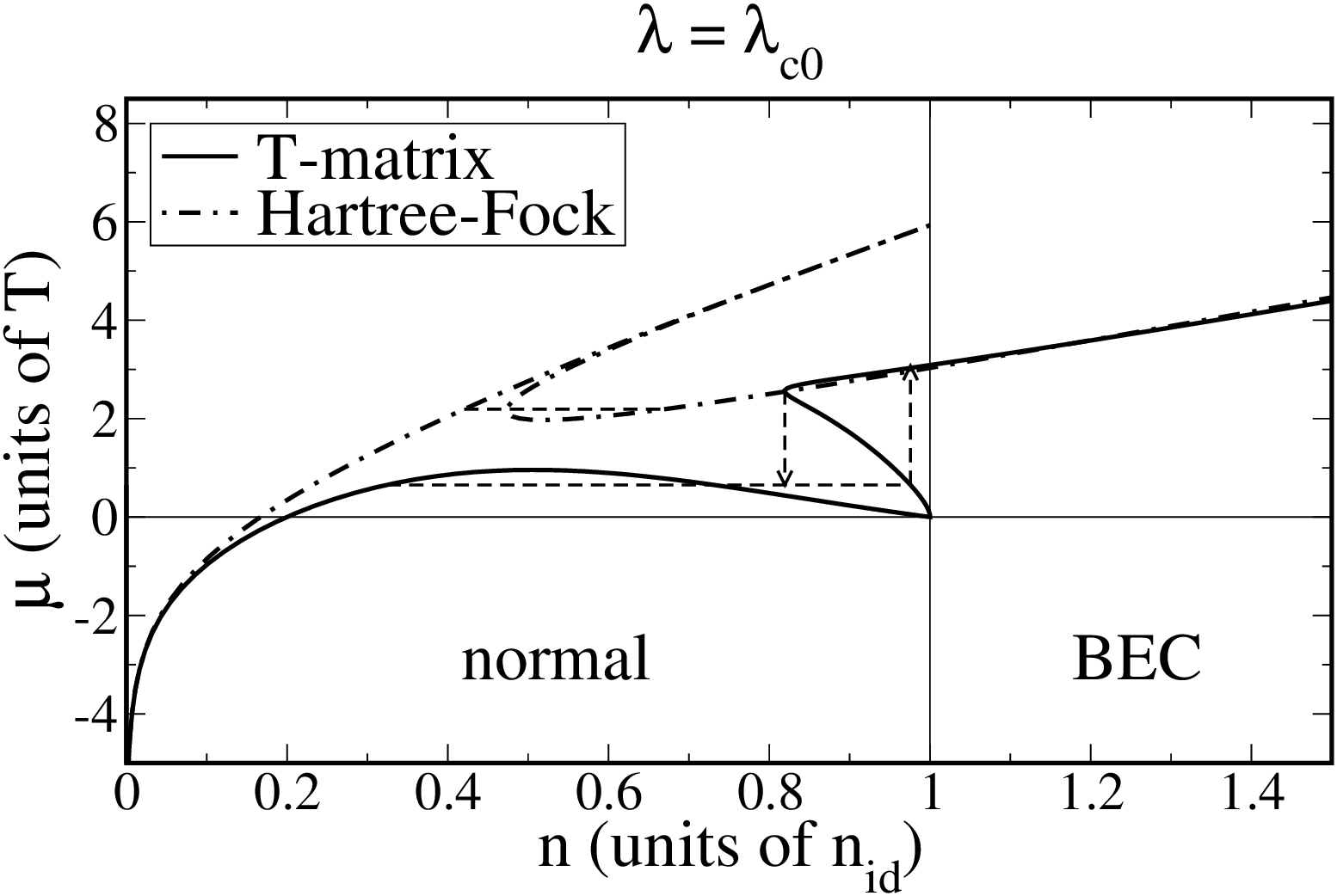}
\caption{Chemical potential in 
Hartree-Fock and T-matrix approximation for strong repulsive interaction, the horizontal broken lines correspond to the Maxwell construction, the vertical arrows mark the density hysteresis.
\label{cnt6}}
\end{figure}

Fig~\ref{cnt6} shows the chemical potential
in the two approximations for a stronger repulsion. In 
Hartree-Fock approximation, i.e., the
dashed-dotted line, there is no qualitative change. However, for the T-matrix
approximation (full line) a multivalued region appears for \mbox{$\lambda>0.23$~$\lambda_{\rm c0}$}, which cannot be avoided
by the Maxwell construction. Therefore we attribute a true physical relevance
to this behavior and interpret it as appearance of a hysteresis. 
Reaching the end of the coexistence region at \mbox{$n\approx 0.98$~$n_{\rm id}$} from below the chemical potential jumps from $0.65$~$T$ to $3.07$~$T$. Decreasing the density the
chemical potential decreases and jumps back to $0.65$~$T$ at a smaller density near
$0.82$~$n_{\rm id}$. This can be understood as hysteresis behavior. 

The alternative view is to consider the density as function of the chemical potential. Normally adding a particle costs energy due to repulsion. In the multivalued region we have the situation that with increasing chemical potential the density drops. This indicates strong rearrangement and correlations which make the effective interaction attractive. 
Therefore a hysteresis appears due to the strong correlation.

\subsection{Condensate density}

\begin{figure}[t]
\includegraphics[width=8cm]{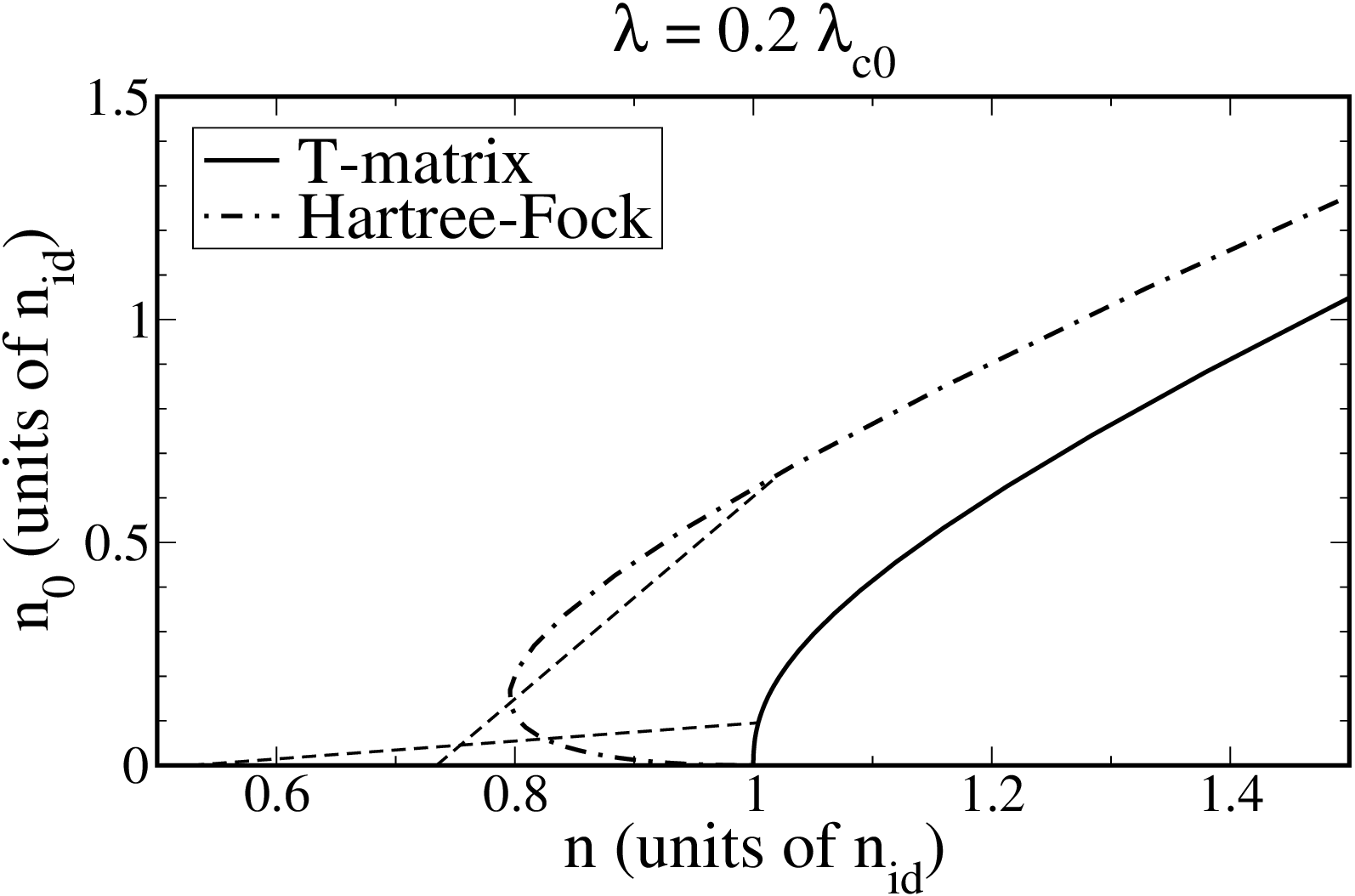}
\caption{Condensate density in 
Hartree-Fock and T-matrix approximation for weak repulsive interaction, the broken lines correspond to the construction according to equation (\ref{n0lin}).
\label{n0nth}}
\end{figure}

\begin{figure}[t]
\includegraphics[width=8cm]{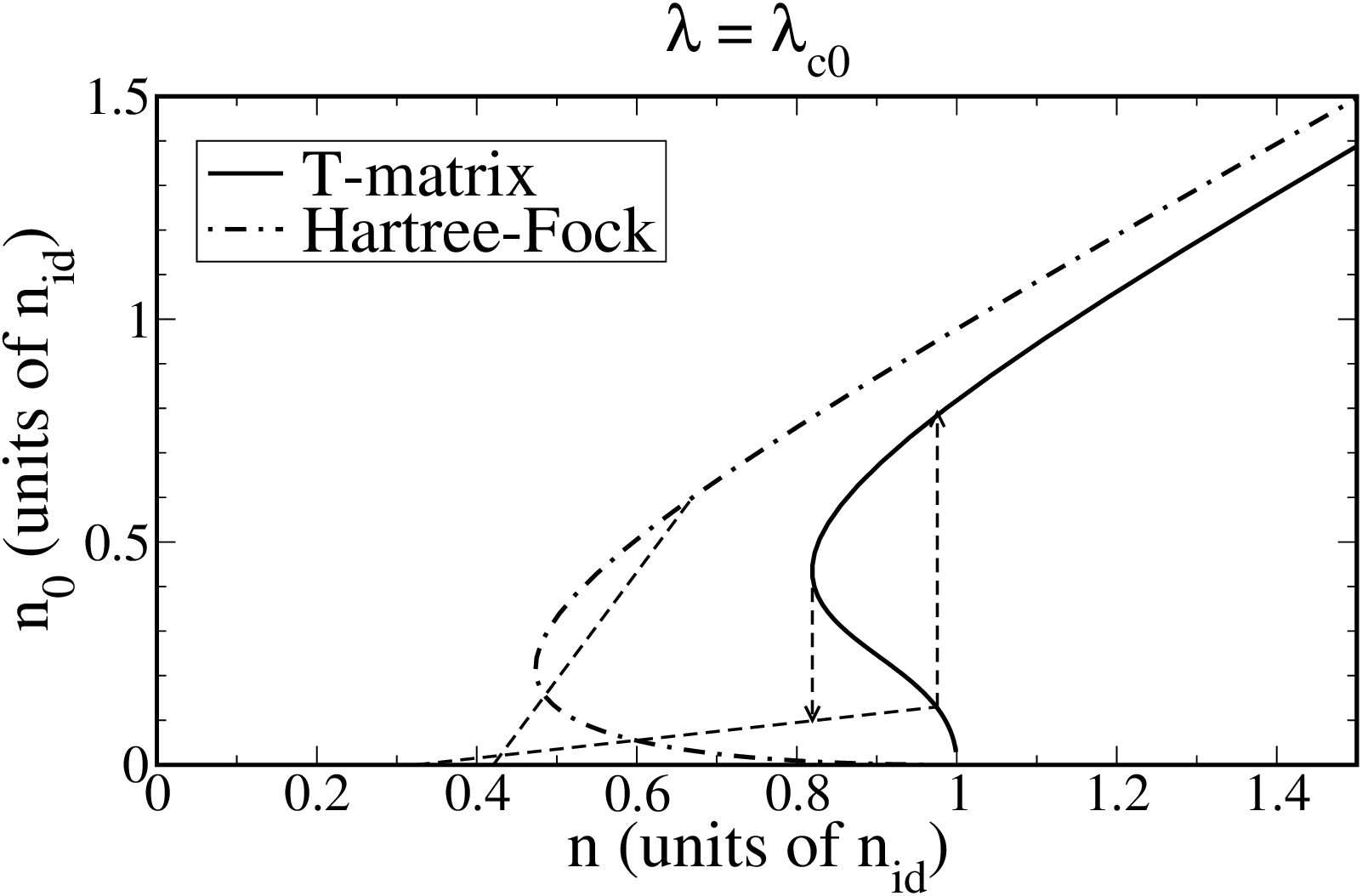}
\caption{Condensate density in 
Hartree-Fock and T-matrix approximation for strong repulsive interaction, the broken lines corresponds to the construction according to equation (\ref{n0lin}), the vertical arrows mark the density hysteresis.
\label{n0n2}}
\end{figure}

The dependence of the condensate density on the total one for weak and strong interaction are illustrated in Figs.~\ref{n0nth} and \ref{n0n2}. 
The condensate density in 
Hartree-Fock approximation shows an unphysical multivalued region as well. This behavior has already been found by Huang et al. \cite{HYL57}. They also proposed a solution to the problem, i.e., taking into account the first-order phase-transition. 

As already mentioned there is a coexistence of a high- and a low-density phase during the first-order phase-transition. The corresponding properties are labeled with subscript h and l in the following. The phases are separated, therefore their volumes add to the total one $\Omega=\Omega_{\rm l}+\Omega_{\rm h}$, which shall be fixed. The gas is driven through the phase transition by increasing the total number of particles $N=N_{\rm l}+N_{\rm h}$. The phase transition takes place in the region of total density $n_1\le n=N/\Omega\le n_2$. As intensive parameter the density within the low-density phase $N_{\rm l}/\Omega_{\rm l}$ is constant during the phase transition. At the lower border of the phase transition all particles are in the low-density phase, i.e., $N_{\rm l}/\Omega_{\rm l}=n_1$. Analogously the density within the high-density phase equals the total one at the upper border of the phase transition $N_{\rm h}/\Omega_{\rm h}=n_2$. With these conditions one can find the density of the high-density phase within the total volume
\be
\frac{N_{\rm h}}{\Omega}=n_2 \frac{n-n_1}{n_2-n_1}.
\ee
This leverage relationship shows that $N_{\rm h}/\Omega$ changes linearly with the total density $n$. An analogous relationship can also be obtained for $N_{\rm l}/\Omega$ \cite{N12}. In the present case a certain fraction of the high-density phase forms a BEC. As the temperature and the density $N_{\rm h}/\Omega_{\rm h}$ are constant during the phase transition
also the condensate fraction $N_{\bf 0}/N_{\rm h}$ is constant and equal to its value at the upper border of the phase transition, i.e., $N_{\bf 0}/N_{\rm h}=n_{\bf 0}(n_2)/n_2$. The condensate density within the total volume is therefore
\be
n_{\bf 0}(n)=\frac{N_{\bf 0}}{\Omega}=n_{\bf 0}(n_2) \frac{n-n_1}{n_2-n_1}.
\label{n0lin}
\ee
According to this equation BEC starts already at the lower border of the phase transition $n_1$, which is always smaller than the ideal critical density $n_{\rm id}$, and the condensate density increases linearly with the total one $n$ during the phase transition. This linear construction according to equation (\ref{n0lin}) is illustrated in Figs.~\ref{n0nth} and \ref{n0n2} as broken lines. The borders $n_1$ and $n_2$ of the first-order phase-transition have to be calculated from the Maxwell construction of the chemical potential or pressure.

\begin{figure}[t]
\includegraphics[width=8cm]{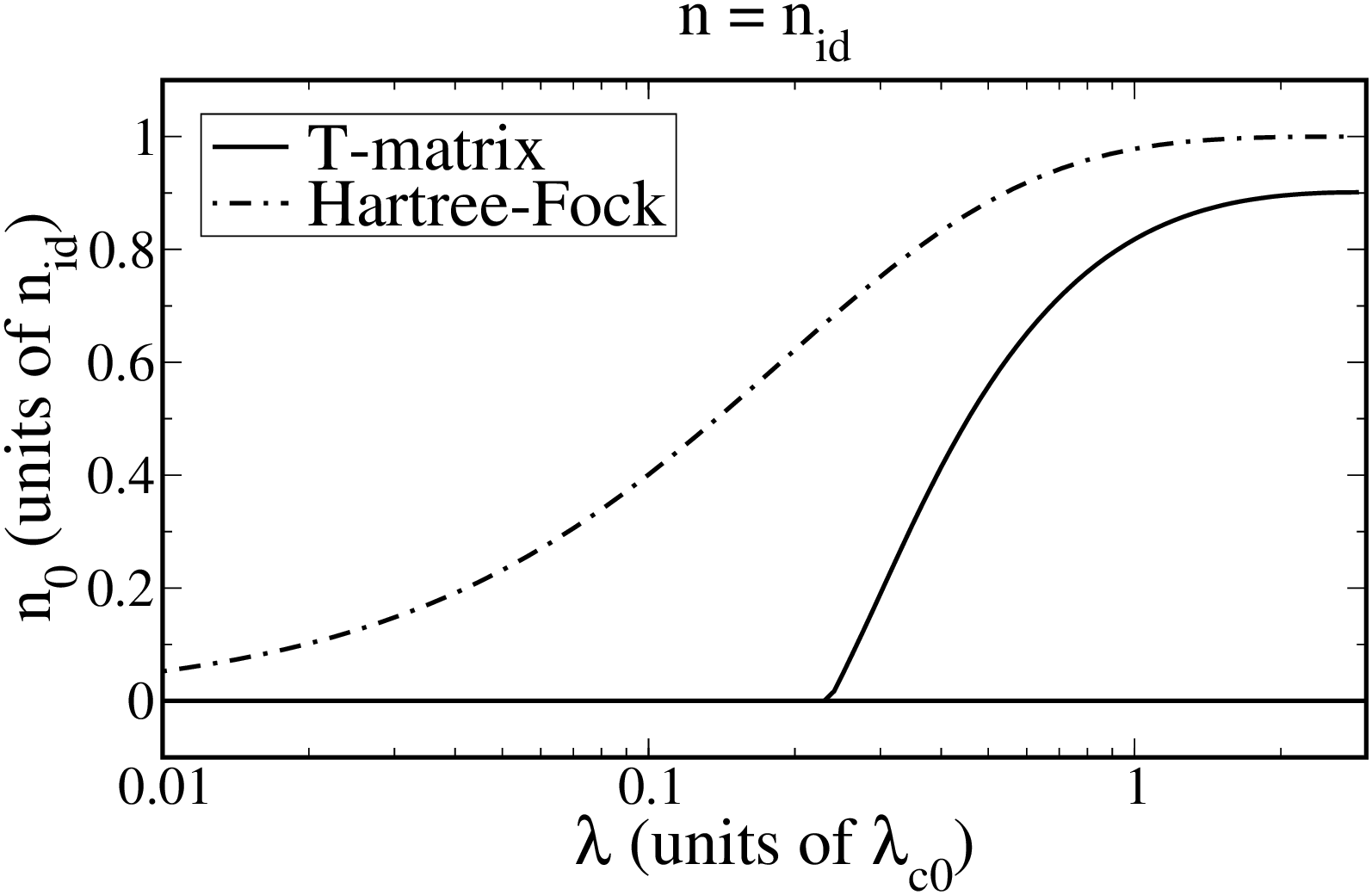}
\caption{Solutions for the condensate density at the critical point in 
Hartree-Fock and T-matrix approximation.
\label{n02l2}}
\end{figure}

With this linear construction the multivalued region for the 
Hartree-Fock approximation can be avoided. This construction is also possible for the \mbox{T-matrix} approximation, where we observe a surviving of the multivalued region for stronger interaction and a hysteresis like in the chemical potential. The existence of the multivalued region is illustrated in Fig.~\ref{n02l2} in terms of the condensate density without linear construction. Together with the unphysical region there appears a second finite solution for the condensate density at the critical point $n=n_{\rm id}$. 
Besides the trivial solution $n_{\bf 0}=0$ there is always a second finite solution for the 
Hartree-Fock approximation (dashed-dotted line) while for the T-matrix approximation (solid line) the second solution appears only above some critical interaction strength of about $0.23$~$\lambda_{\rm c0}$.

\section{Comparison with experiment and other models}
\label{sectn4}
A direct comparison of the calculations presented here with experiments is limited by several reasons. First the atoms considered in this paper have zero total spin, while the alkali atoms in the experiments have finite total spin. Second the atomic gas in the experiment is trapped magneto-optically or in an optical lattice \cite{MO06} and therefore not homogeneous, in contrast to the gas considered here. And third the T-matrix approximation presented here is not sufficient to describe the shift of the critical temperature \cite{BBHLV01}, which is however measurable in experiment and has also an influence on the behavior of the condensate \cite{SCTH11,STCHH11}. One can use inverse expansions to extend the T-matrix to describe such temperature shift \cite{MMS07}.

For $^4$He the parameters of the Yamaguchi interaction can be fit to the s-wave scattering length $a_0=93$~{\AA} and effective range $r_0=7.298$~{\AA} \cite{JA95}, yielding $\gamma=0.015$~{\AA}$^{-1}$ and $\lambda=2.24$~$\lambda_{\rm c0}\cdot2\sqrt{\pi}/\gamma\Lambda_{\rm dB}$. The temperature corresponding to $\gamma=2\sqrt{\pi}/\Lambda_{\rm dB}$ as chosen for all plots is $T=1.4$~mK$\cdot k_{\rm B}$. However only the curves for the \mbox{T-matrix} approximation depend on $\gamma$ and $T$. The curves for the other approximations are independent of $\gamma$ and therefore independent of $T$ due to the scaling.

\begin{figure}[t]
\includegraphics[width=8cm]{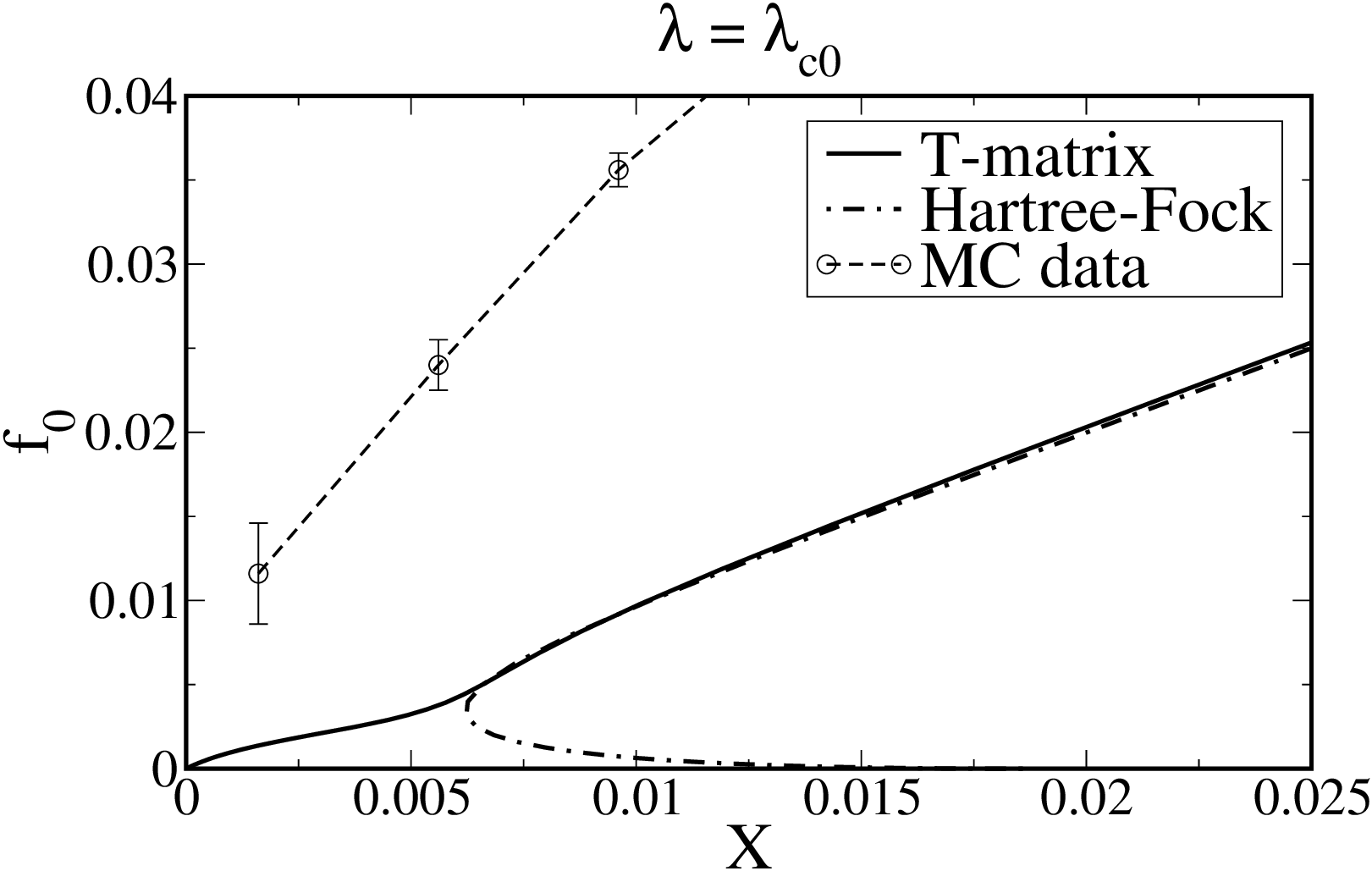}
\caption{Dimensionless plot of the condensate density (\ref{17}) depending on the chemical potential (\ref{18}), in comparison with MC data from Ref.~\cite{PRS04}.
\label{f0X}}
\end{figure}

A comparison of our results with a Monte Carlo (MC) calculation \cite{PRS04} of the dimensionless condensate density
\be
f_{\bf 0}=\frac{n_{\bf 0}\hbar^6}{m^3T^2\lambda}=\frac{n_{\bf 0}}{n_{\rm id}}\frac{\lambda_{\rm c0}}{\lambda}\frac{2.61}{16\sqrt{\pi}^5}
\label{17}
\ee
depending on the dimensionless chemical potential
\be
X=\frac{\mu\hbar^6}{m^3T^2\lambda^2}=\frac{\mu}{T}\left(\frac{\lambda_{\rm c0}}{\lambda}\right)^2\frac{1}{32\pi^2}
\label{18}
\ee
yields qualitative agreement. The result is shown in Fig.~\ref{f0X}. With the higher order \mbox{T-matrix} approximation (solid line) the back bending of the Hartree-Fock approximation (dashed dotted line) can be removed. However there is no quantitative agreement of the \mbox{T-matrix} approximation with the MC data (broken line).

We find that for a repulsive Bose gas an additional first-order phase-transition is accompanying the BEC. Although the onset of BEC is changed by the first-order phase-transition the BEC transition itself remains continuous, i.e., except for the hysteresis there is no jump in the condensate density.  
Nevertheless, this is probably an artifact of the approximations used, since the general believe is, that for such a system there is only the continuous BEC phase transition \cite{HB03,PRS04}.

For the case of strong correlations we find a region of decreasing density with increasing chemical potential which cannot be removed by the Maxwell construction. From the viewpoint of experimental realization one could remove particles from the system to reach this region. Then the chemical potential will drop at a specific density to a lower value. 
We suggest that this indicates an onset of a rearrangement phase transition which shows up at higher correlations as hysteresis. 

\section{Summary and conclusions}
\label{sectn5}

For a repulsive Bose gas the continuous BEC transition is closely related to a first-order phase-transition. The instability causing the first-order phase-transition appears due to the onset of BEC. In view of the first-order phase-transition, BEC sets in already at a lower density than initially expected. In spite of the repulsive interaction, the instability of the system is caused by the attraction in momentum space and medium effects closely related to the bosonic character of the particles. BEC sets in with the first-order phase-transition and the condensate density increases linearly during this phase-transition. 
The physical
relevance is justified by the successive higher level of approximation used
here. Lower-level approximations show artificial multivalued regions which
can be avoided by the Maxwell construction. With higher level of approximations
this multivalued region 
shrinks and for weak interaction vanishes for the T-matrix
approximation. In the case of strong
interactions we observe that besides the first-order phase-transition region, multiple
solutions appear for the T-matrix
approximation which are interpreted as density hysteresis.

\end{document}